# On the velocity autocorrelation function of a Brownian particle

Roumen Tsekov and Boryan Radoev
Department of Physical Chemistry, University of Sofia, 1164 Sofia, Bulgaria

Memory effect of Brownian motion in an incompressible fluid is studied. The reasoning is based on the Mori-Zwanzig formalism and a new formulation of the Langevin force as a result of collisions between an effective and the Brownian particles. Thus, the stochastic force autocorrelation function with finite dispersion and the corresponding Brownian particle velocity autocorrelation function are obtained.

It has been theoretically accepted by Langevin [1] that the interaction of a macroparticle with a medium can be presented by two forces: a friction force, related to the concept of fluid hydrodynamic viscous resistance [2], and a random force, called usually the Langevin force, which reflects the stochastic character of collisions. According to this picture, the momentum balance of the Brownian motion has the form of a generalized Langevin equation [3-5]

$$MdU/dt + \hat{G}U = F \qquad (1)$$

where $M$ and $U$ are the mass and velocity of the Brownian particle, $\hat{G}U$ is the friction force and $F$ is the Langevin force with a zero mean value $<F>=0$. The Langevin force is not correlated with the particle velocity in previous moments, $<F(t+\tau)U(t)>=0$ for $\tau>0$. Due to the common molecular-kinetic origin of the two forces $\hat{G}U$ and $F$ they are not independent. Their link, named by Kubo [3] the second fluctuation-dissipation theorem, is most generally formulated by Mori [4] and Zwanzig [6]

$$k_B T \hat{G} U(t) = \int_0^t \mathbb{K}_{FF}(t-s) \cdot U(s) ds \qquad (2)$$

where $\mathbb{K}_{FF}(\tau) \equiv <F(t+\tau)F(t)>$ is the autocorrelation function of the Langevin force and $T$ is temperature. The operator $\hat{G}$ is proportional to the friction coefficient $b$ and their relationship, expressed by Eq. (2), is given by the well-known equation [3, 5, 7]

$$k_B T b \mathbb{I} = \int_0^\infty \mathbb{K}_{FF}(\tau) d\tau \qquad (3)$$

where $\mathbb{I}$ is the unit tensor. At slow translation of a spherical particle in an incompressible fluid (Stokes flow [2]) the friction coefficient $b$ is equal to $6\pi\eta R$ with $\eta$ being the fluid viscosity and $R$ being the particle radius. An equation for the evolution of the Brownian particle velocity autocorrelation function $\mathbb{K}_{UU}(\tau) \equiv <U(t+\tau)U(t)>$ can be obtained from Eqs. (1) and (2) [5, 8]

$$k_B T M d\mathbb{K}_{UU}/d\tau + \int_0^\tau \mathbb{K}_{FF}(\tau-s)\cdot\mathbb{K}_{UU}(s)ds = 0$$

Further the Laplace transformation of time-dependent functions will be employed and Laplace images will be denoted by a tilde superscript. Using the Maxwell expression for the dispersion $\mathbb{K}_{UU}(0) = k_B T \mathbb{I}/M$ of the Brownian particle velocity, the equation above acquires the form

$$(k_B T M \mathbb{I} p + \tilde{\mathbb{K}}_{FF})\cdot\tilde{\mathbb{K}}_{UU} = (k_B T)^2 \mathbb{I} \qquad (4)$$

where $p$ is the Laplace transformation variable.

Usually the friction force $\hat{G}U$ is approximated by $bU$, the Langevin force autocorrelation function tends to $2k_B T b\mathbb{I}\delta(\tau)$, and Eqs. (1) and (4) acquires their classical forms [3, 5, 7]

$$MdU/dt + bU = F \qquad \tilde{\mathbb{K}}_{UU} = k_B T\mathbb{I}/(Mp+b) \qquad \mathbb{K}_{UU} = (k_B T/M)\mathbb{I}\exp(-b\tau/M)$$

In the traditional mechanical treatments [5, 7] the transition $\hat{G}U \to bU$ is analyzed on a microscopic level as a function of the ratio $m/M$ between the masses of the fluid and Brownian particles. It is shown that in the mass point approximation $\hat{G}(m/M \to 0)U = bU$. For objects with a finite sufficiently large size and macroscopic interaction with the surrounding medium the magnitude of the mass $m$ should be determined by the mass of fluid displaced by the Brownian particle, i.e. $m = \pi R^3 \rho$ [9-11] with $\rho$ being the fluid mass density. This can be shown [10, 11] in the classical result for the frequency dependent spectral density of the Langevin force autocorrelation function

$$\tilde{\mathbb{K}}_{FF} = k_B T b\mathbb{I}(1+\sqrt{pR^2\rho/\eta}) \qquad (5)$$

The limit, at which $\tilde{\mathbb{K}}_{FF}(p\tau_r)$ ($\tau_r = M/b$ is the relaxation time of the Brownian particle velocity) transforms into a white noise, corresponds to the limit $R^2\rho/\eta\tau_r \sim \pi R^3\rho/M \to 0$.

The aim of the present paper is to describe in more details the behavior of the Langevin force autocorrelation function as compared to the classical models, which show some physical discrepancies. For instance, the dispersion of the Langevin force corresponding to the spectral density (5) is an infinite quantity, which is unphysical. The present new model is based on the assumption that the action of the surrounding medium can be treated as an impact of an effective particle on the Brownian particle, which allows a useful reformulation of the Langevin force. The interaction between a Brownian particle and the surrounding medium can be generally presented by an operator $\hat{C}(V,U)$ acting of the velocity $V$ of an effective particle and the velocity $U$ of the Brownian particle. Since the interaction force should be independent of the whole system drift velocity, it follows that $\hat{C}(V,U) = \hat{C}(V-U)$. In accordance with the classical theory of collisions [7], the operator $\hat{C}$ is a linear one and hence the momentum balance of the Brownian particle acquires the form

$$MdU/dt = \hat{C}V - \hat{C}U$$

This result is equivalent of Eq. (1) and assuming identity of the operators $\hat{C}$ and $\hat{G}$ it follows an expression for the Langevin force

$$F(t) - F(0) = \hat{G}V = \int_0^t \mathbb{K}_{FF}(t-s) \cdot V(s)ds / k_B T \tag{6}$$

Using the stationary nature of the considered processes, $<V> = 0$, Eq. (6) and its obvious consequence $k_B T(dF/dt)(0) = \mathbb{K}_{FF}(0) \cdot V(0)$ yields a link between the autocorrelation functions of the Langevin force and of the velocity of the effective particle $\mathbb{K}_{VV}(\tau) \equiv <V(t+\tau)V(t)>$

$$-(k_B T)^2 d\mathbb{K}_{FF}/d\tau = \mathbb{K}_{FF}(0) \cdot \int_0^\tau \mathbb{K}_{VV}(s) \cdot \mathbb{K}_{FF}(\tau-s)ds$$

which, after application of the Laplace transformation, acquires the form

$$(k_B T)^2 [\mathbb{K}_{FF}(0) - p\tilde{\mathbb{K}}_{FF}] = \mathbb{K}_{FF}(0) \cdot \tilde{\mathbb{K}}_{VV} \cdot \tilde{\mathbb{K}}_{FF} \tag{7}$$

The employment of this result requires an adequate expression for the autocorrelation function of the velocity $V$ by characteristics of the medium. Because the velocity field excited by the Brownian particle in an infinite incompressible fluid has the same shape and size of the macroparticle (the mean depth $4\pi R^2\eta/b$ of the hydrodynamic field [2] is of order of $R$) the effective particle will have the same geometric parameters as those of the Brownian particle. Of course, the mass of the effective particle will be equal to $m = \pi R^3\rho$. Thus the velocity autocorrelation function of the effective particle can be expressed by the Brownian particle velocity autocorrelation function from Eq. (4) as follows

$$\tilde{\mathbb{K}}_{VV} = \tilde{\mathbb{K}}_{UU}(M=m) = (k_BT)^2(k_BTm\mathbb{I}p + \tilde{\mathbb{K}}_{FF})^{-1} \tag{8}$$

It is assumed here that $\tilde{\mathbb{K}}_{FF}$ is independent on the Brownian particle mass, which follows from the independence of the Langevin force on $M$; $\tilde{\mathbb{K}}_{FF}$ from Eq. (5) has the same behavior.

An explicit expression of the Langevin force autocorrelation function follows from the results above. Substituting $\tilde{\mathbb{K}}_{VV}$ from Eq. (8) into Eq. (7) leads to the following expression

$$\tilde{\mathbb{K}}_{FF} = k_BTb\mathbb{I}[\sqrt{1+(p\tau_m/2)^2} - p\tau_m/2] \tag{9}$$

where $\tau_m = m/b$ is the correlation or memory time. The inverse Laplace transformation of Eq. (9) provides an analytical expression for the Langevin force autocorrelation function

$$\mathbb{K}_{FF} = k_BTb\mathbb{I}J_1(2\tau/\tau_m)/\tau \tag{10}$$

where $J_1$ is the Bessel function of first kind and first order. According to Eq. (10) the modulus of this autocorrelation function tends to zero at large times as $\tau^{-3/2}$, thus leading to a known result [9, 10]. Equations (4) and (9) allow the obtaining of the spectral density of the autocorrelation function of the Brownian particle velocity in the form

$$\tilde{\mathbb{K}}_{UU} = k_BT\mathbb{I}/b[\sqrt{1+(p\tau_m/2)^2} + p(\tau_r - \tau_m/2)] \tag{11}$$

where $\tau_r = M/b$ is the Brownian particle relaxation time. In some particular cases the inverse Laplace transformation of Eq. (11) is possible to be performed and the corresponding solutions are given below:

$\mathbb{K}_{UU}(\tau_m \ll \tau_r) = (k_BT/M)\mathbb{I}\exp(-\tau/\tau_r)$      heavy Brownian particle

$\mathbb{K}_{UU}(\tau_m = \tau_r) = (k_BT/M)\mathbb{I}J_1(2\tau/\tau_r)(\tau_r/\tau)$      neutral, equal densities

$\mathbb{K}_{UU}(\tau_m = 2\tau_r) = (k_BT/M)\mathbb{I}J_0(\tau/\tau_r)$      lighter Brownian particle

where the second expression has the same time-dependence as $\mathbb{K}_{FF}$ from Eq. (10). It should be noted that such oscillatory-decaying velocity autocorrelation functions are observed in numerical simulations [8, 12]. Equation (11) shows that different behavior of $\mathbb{K}_{UU}$ will be observed at small and large times, which has been discovered by numerical simulations [13, 14] as well (the so-called long-time tails). Also by numerical simulations [12] the dependence of $\mathbb{K}_{UU}$ by the ratio between the fluid and Brownian particle densities is obtained, which is similar to the predicted by Eq. (11) dependence of $\mathbb{K}_{UU}$ on the corresponding ratio $\tau_m / \tau_r = m/M$.

In the frames of the linear hydrodynamics the problem of finding of the Langevin force autocorrelation function can be solved exactly and the result is Eq. (5). As noted before it diverges at $\tau = 0$. This inconsistency is inherent for the macroscopic description of diffusive processes and its elimination requires employment of microscopic concepts. The present modeling of the interaction of the Brownian particle with the surrounding medium as impacts by an effective particle is not directly related to the linear hydrodynamics. Hydrodynamic concepts are used only for determination of the parameters of the effective particle and the autocorrelation function $\mathbb{K}_{FF}$. For instance, the size of the effective particle is assumed to be proportional the Brownian particle size in analogy with the particle-fluid viscous interaction in a Stokes flow. The obtained in Eq. (10) finite dispersion $\mathbb{K}_{FF}(0) = k_B T b \mathbb{I} / \tau_m$ of the Langevin force possesses a clear physical explanation. The velocity dispersion $\mathbb{K}_{FF}(0) \tau_m^2 / m^2$, generated by the Langevin force, corresponds to the classical Maxwell expression $<VV> = k_B T \mathbb{I} / m$.

[1]  P. Langevin, *Compt. Rend. Acad. Sci. (Paris)* **146** (1908) 530
[2]  L.D. Landau and E.M. Lifshitz, *Fluid Mechanics*, Pergamon, New York, 1959
[3]  R. Kubo, *Rep. Prog. Phys.* **29** (1966) 255
[4]  H. Mori, *Prog. Theor. Phys.* **33** (1965) 423
[5]  D. Forster, *Hydrodynamic Fluctuations, Broken Symmetry and Correlation Functions*, Benjamin, Massachusetts, 1975
[6]  R. Zwanzig, *Annu. Rev. Phys. Chem.* **16** (1965) 67
[7]  P. Resibois and M. de Leener, *Classical Kinetic Theory of Fluids*, Wiley, New York, 1977
[8]  B.J. Berne and G.D. Harp, *Adv. Chem. Phys.* **17** (1970) 63
[9]  R. Zwanzig and M. Bixon, *Phys. Rev. A* **2** (1970) 2005
[10] T.S. Chow and J.J. Hermans, *J. Chem. Phys.* **56** (1972) 3150
[11] D. Bedeaux and P. Mazur, *Physica A* **76** (1974) 247
[12] B.J. Alder, D.M. Gass and T.E. Wainwright, *J. Chem. Phys.* **53** (1970) 3813
[13] B.J. Alder and T.E. Wainwright, *Phys. Rev. A* **1** (1970) 18
[14] G. Subramanian, D.G. Levitt and H.T. Davis, *J. Chem. Phys.* **60** (1974) 591